# An efficient way to manage ranges of data with Wise Red-Black Trees

Alberto Boffi[1]

Department of Electronics, Information and Bioengineering - Politecnico di Milano, Italy

**Abstract**: This paper describes the most efficient way to manage operations on ranges of elements within an ordered set. The goal is to improve existing solutions, by optimizing the average-case time complexity and getting rid of heavy multiplicative constants in the worst-case, without sacrificing space complexity. This is a high-impact operation in practical applications, performed by introducing a new data structure called Wise Red-Black Tree, an augmented version of the Red-Black Tree.

# 1 Introduction

## 1.1 The problem

Let's analyze a data management method common in several fields, such as logistics.
Consider an ordered set: instead of working on single elements, we want to do multiple operations on consecutive elements, i.e. operate on what I'll call "blocks" of elements.

To clear things up, consider the following example:

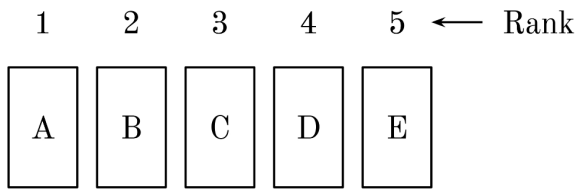

We want to access a (possibly unary) block of elements. The way we are going to identify a sequence of elements will be through the rank of the first one, and the total number of elements. For example, we want to change three items starting from the second one, updating their content with *["F", "G", "H"]:*

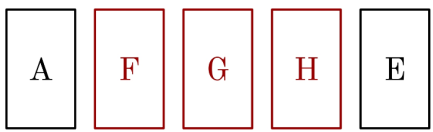

---

[1] E-mail: alberto2(*dot*)boffi(*at*)mail(*dot*)polimi(*dot*)it

We now want to delete the same block. The set becomes:

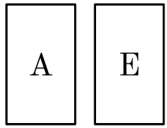

Also regarding the append, we are interested in inserting one block at a time:

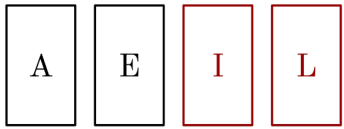

## 1.2 Standard solutions

Let *n* be the number of elements in the structure.

Although arrays allow direct access, using them to represent this type of set would cause efficiency problems. They in fact have static size, and periodic entries would therefore require a compromise between an excessive use of memory, and continuous reallocations of time complexity $T(n) = \Theta(n)$. Deletion also causes problems, 'cause all the elements following the deleted ones should be shifted. Even this operation, in the worst-case has a cost of $T(n) = \Theta(n)$.

Even a list, although it has dynamic size and does not need any shift after a deletion, requires for each access a complexity $T(n) = O(n)$.

It's anyway known that the order relationship between elements prompts us to manage them in a *tree*. In particular, given the operations we want to perform, we need a *binary search tree*.
Let's consider a self-balancing binary search tree, such as a Red-Black Tree.
Given a block of $m$ elements, a search or an insertion will cost $T(n) = O(m \cdot log(n))$[a]. In the deletion, if the rank of each node is stored in its key, after removing an element we have to shift keys of subsequent ones: the overall time required is $T(n) = \Theta(n + m \cdot log(n))$.

So we fall into an *order-statistic tree*. We can in this way access the k-th smallest element efficiently, avoiding shifting keys after a deletion and getting $T(n) = O(m \cdot log(n))$ for every operation, both in worst and average-case[a].

## 1.3 Toward a new solution

Time required isn't so bad, but note: despite classical binary search trees exploit the order relationship present in the set, they do not exploit the information on the sequentiality of the elements they treat.

So I wondered if it was possible to take advantage of the room for a further efficiency improvement, offered by the management of the whole block we have to deal with as a single unit, rather than operate on each individual node independently.

The result is a great optimization of the average-case, and an erasing of significant multiplicative constants that influence the worst-case too. All without sacrificing space complexity.
This positively affects practical applications, working both on large and small cardinality sets.

I'll show you how this is possible by presenting a new augmented Red-Black Tree, slightly different from classic order-statistic trees, that I decided to call **Wise Red-Black Tree**.

# 2 Pseudocode conventions

To access a field of a structure *struct*, I will use the notation *struct.field*.
In order to obtain a more intuitive code, I will consider each structure as an implicit pointer.
A pointer not referring to any object has the value *NIL*.

I'm then considering the default implementation of the tree, which requires to store in each node the reference *p* to the parent.

In this paper, pseudocode will just be used at the beginning to fix some basic concepts. The full implementation of a Wise Red-Black Tree, written in C language, is available at https://github.com/albertoboffi/wise-red-black-tree.

# 3 A first look at Wise Red-Black Trees

Let's first see the basics behind Wise Red-Black Trees.

## 3.1 Node structure

Unlike order-statistic trees, the additional field (*leftSize*) stored in each node of a Wise Red-Black Tree indicates only the size of the *left* subtree, *excluding* sentinel NILs.

```
node{
  key,
  color,
  left,
  right,
  p,
  leftSize,

  data
}
```

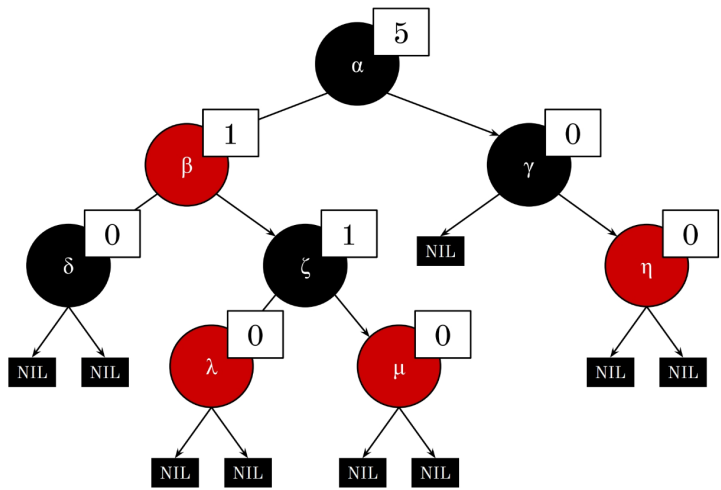

Figure 3.1.1: Values of leftSize fields in a sample tree

In addition to the references to the root and sentinel NILs, it will be kept a reference to the node with maximum key, obviously initialized to *NIL*.

## 3.2 Find k-th smallest element

The concept describing the search for the k-th smallest element is not affected.

```
kthSmallest(R, k){
  pos = R.leftSize + 1
  if (k == pos) return R
  if (k < pos) return kthSmallest(R.left, k)
  return kthSmallest(R.right, k-pos)
}
```

### Time complexity

Both average and worst-case complexity are proportional to the height of the tree, which in a self-balancing tree is logarithmic with respect to the number of nodes: $T(n) = O(log(n))$[a].

## 3.3 Rotation

After a rotation in a wise Red-Black tree, we need to make sure that each *leftSize* field contains the correct value.

To do so, it's sufficient to observe that a rotation reverses the genealogy of only two nodes.

Let's first analyze a left rotation of a node *X*. Consider *Y*=*X*.*right* <u>before</u> the rotation. After the rotation, *Y* will have additional nodes in the left subtree, consisting of *X* and the left subtree of *X*.

Consider instead a right rotation of X, with Y=X.*left before* the rotation. After the rotation, in the left subtree, X will no longer have Y and the left subtree of Y.

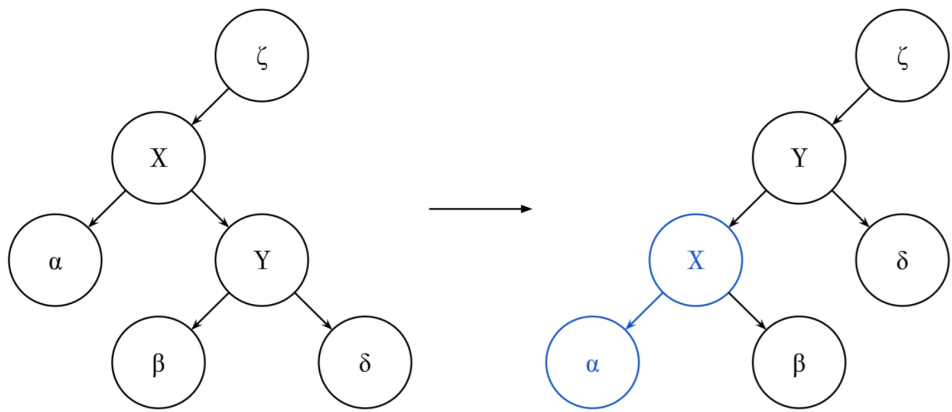

Figure 3.3.1: Left rotation

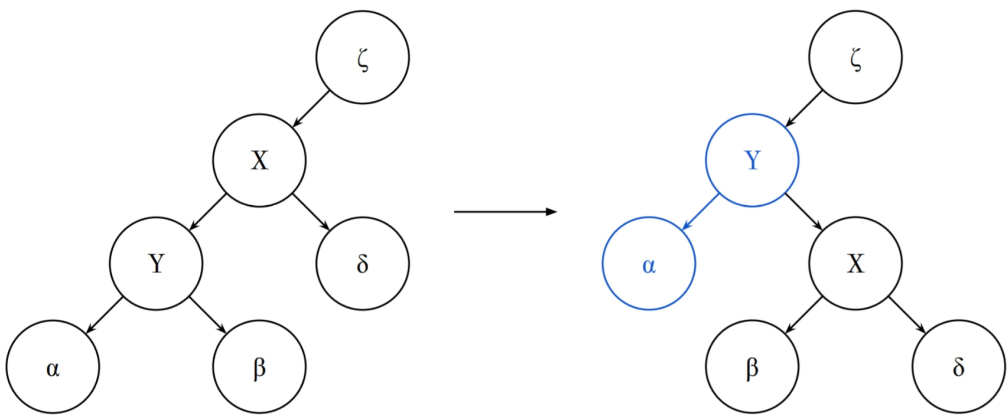

Figure 3.3.2: Right rotation

So, only a small adjustment is required in both types of rotation:

```
leftRotate(X){                          rightRotate(X){
   …                                       …
   Y.leftSize += (X.leftSize + 1)          X.leftSize -= (Y.leftSize + 1)
}                                       }
```

## Time complexity

The adjustment requires only a single instruction, leaving the complexity unchanged at $T(n) = O(1)$.

# 4 Operations of interest

Once the preliminaries have been completed, let's see how to perform the operations we're interested in, involving blocks of elements.

## 4.1 Search

We want to do a generic access to a block of elements, for example to modify the content or print it on screen.
As said before, to locate the block we need its size *m* and the rank *k* of the first element.

To access the first element, we will look for the k-th smallest node of the tree.
At this point, iteratively, given a node, to search for the next element of the block we will search for the inOrder successor. We will proceed this way until the end of the block.

### Time complexity

The search for the k-th smallest element has complexity $T(n) = O(log(n))$[a].

Given the size *m* of the block, after this we have *m-1* sequential searches of the InOrder successor: each successor found represents the predecessor of the following node to search.
Although finding a single successor costs $T(n) = O(log(n))$, the total complexity of *m-1* consecutive searches is *not* given by the sum of the complexities of the individual searches, but by $T(n) = O(m - 1 + log(n)) = O(m + log(n))$[a].

The total temporal complexity is consequently
$T(n) = O(m + 2log(n)) = O(m + log(n)) = O(max(m, log(n)))$.

## 4.2 Insertion

### 4.2.1 Append

Let's first see how to append a block.

By keeping the *maxNode* reference to the maximum key node, it's not required any preliminary search before the append. For each element of the block, we just have to enter a node to the right of *maxNode*: to represent the rank of the new element, the key of the node will be equal to *maxNode.key+1*, while the *leftSize* field will be set to 0.
The rest of the algorithm takes place as in a normal Red-Black Tree, coloring the node red and restructuring to compensate for violations.
We can now update *maxNode* by making it reference to the newly inserted node, and move one to the next element.

If the tree is empty before the append, the only difference is that the first node has to be inserted as the root, with a key 1.

### Time complexity

Since there is no search, the only cost is given by adjustments to rebalance the tree.
In the worst-case, each one of them has a complexity $T(n) = O(log(n))$[a][b], and so we have in total $T(n) = O(m \cdot log(n))$.
In the average-case, fixes cost $T(n) = O(1)$[b], bringing us to $T(n) = O(m)$.

### 4.2.2 Generic insertion

In order to fully show the logic behind insertions in wise Red-Black tree, I'll present the algorithm to operate a generic insertion, in any point of the structure.

I decided not to include at first this operation among the basic ones because, in some cases, it requires extra key management. However, a generic insertion could be required for several features, like for example *undo* a deletion.

I'll suppose we know the key of the nodes to be inserted (this is the scenario of the *undo* of a deletion). If, on the other hand, you can only provide the position *k* where the block has to be inserted, you just have to preliminarily read the key of the *(k-1)-th* and *k-th* smallest elements of the tree. You'll therefore be able to assign a proper key to every element of the block, so that each one is the inOrder successor of the previous[2]. This additional operation will not change the asymptotic complexity.

First of all, as in a classic BST, we have to search for the key of the first node *X* to be inserted, until we come up to a NIL leaf.
In addition, every time we go down to the left of a node *Y* during the descent, we need to increment *Y.leftSize* by the size *m* of the block.
We can now add *X* as in a normal RBT (setting *leftSize* field to 0).
For the following element of the block we don't need any other search: if even after any fix *X* has *NIL* as its right child, the father of the new node to be inserted is *X* itself; otherwise, it'll be *X.right*.

### Time complexity

The search for the position of the insertion has a complexity $T(n) = \Theta(log(n))$ regarding the first node, while $T(n) = \Theta(1)$ for the next ones.
So, again, the cost is dominated by the rebalancing procedure: $T(n) = O(m \cdot log(n))$ in the worst-case, $T(n) = O(m + log(n)) = O(max(m, log(n)))$ in the average-case.

---

[2] "Intermediate" keys can be easily managed passing from natural number domain to real number domain

## 4.2.3 Technical explanation of the algorithm

Management of insertion positions

I'll first show how it's possible, once a node is inserted, to insert the next one in $T(n) = O(1)$. Trivial as it may seem, the entire insertion method is based on the following statement:

**Proposition 4.2.1**
*Let X be a node in a binary search tree. Let's now insert a node Y, such that Y is the inOrder successor of X. Then, after the insertion, Y will be the minimum key node of the right subtree of X.*

**Proof**
*By definition of BST, if X.right≠NIL then the inOrder successor of X is the minimum key node of its right subtree. On the other hand, if X.right=NIL the inOrder successor is the closest progenitor of X, such that X is in its left subtree.*
*Since Y has just been inserted, given the insertion procedure of BST, Y can't have children. This implies that Y is necessarily the minimum key node of the right subtree of X.*

Since each node Y of the block is the inOrder successor of the previous one X, once X is inserted we are in the hypothesis of *Proposition 4.2.1*.
So, if *X.right=NIL*, we can easily add Y as the right child of X.

It's otherwise possible that any fix makes X gain a right child. To understand how this can be possible, we've first to realize that only rotations can change the hierarchy of the tree, so we can limit our study to them.
Let's now analyze the 3 kinds of fixes (I'll reference each case through the enumeration used in [a]):

*Case 1*
Case 1 recursively moves the node to analize up to *X.p.p*. In this scenario, rotations of cases 1 and 2 cannot affect the children of X.

*Case 3*
If *X.p.p.left=X.p* we execute *rightRotate(X.p.p)*.
If *X.p.p.right=X.p* we execute *leftRotate(X.p.p)*.
Again, in both cases right and left subtrees of X are not affected.

*Case 2*
If *X.p.p.left=X.p* we execute *leftRotate(Z)*, where *Z=X.p*. After this, the right child of X will still be *NIL*. So we have to move on to case 3, considering this time Z instead of X. Since at first X has *NIL* as right child, after *rightRotate(Z.p.p)* we have *X.right.left=NIL*.
If *X.p.p.right=X.p* we execute *rightRotate(Z)*, where *Z=X.p*. In this the left child of X will still be *NIL*. Consequently, moving to case 3, after *rightRotate(Z.p.p)* we have again *X.right.left=NIL*.

To conclude, the only way to have *X.right*≠*NIL* after the insertion procedure is that we land *immediately* on case 2. When this happens, we always have *X.right.left*=*NIL*.
For *Proposition 4.2.1*, the next node *Y* must be inserted at the left of *X.right*.

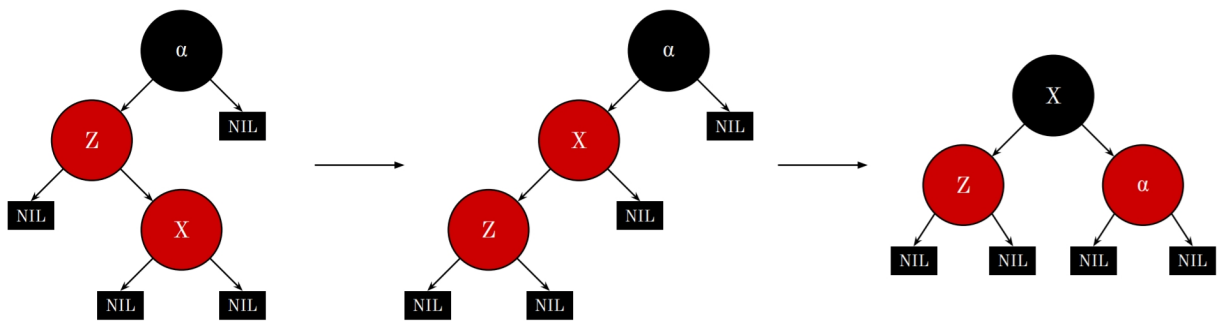

Figure 4.2.1: Node X getting a right child

Management of *leftSize* field

Note that, in order to manage *leftSize* fields, we only have to increase the value by *m* during the first descent, each time we move down to the left.

The all reason lies in the fact that insert each node, make the fixes, and move on to next elements it's equivalent to insert the all block at once, and only then perform the rebalancing procedure on each one.

In fact, imagine inserting *X* and *Y* together, so that we have *Y*=*X.right*, and then eventually compensate for violation on *X*.
As mentioned before, if we ended up in case 1 or 3, *Y* will remain the right child of *X*, while we can observe that in case 2, *Y* will move on the left of *X.right*.
So, in any case, *Y* will land in the same position we deduced considering to fix the tree *before* the insertion.

This important observation allows us, during the first descent, to ideally consider to insert the whole block at once: at the end of the day, all the rotations on each possible restructuring procedure will take care of adjusting the *leftSize* fields.

## 4.3 Deletion

Even to delete a block, we need to know the size *m* of the block and the rank *k* of the first node to be removed.

To access the first element, we have to search for the *k-th* smallest element. This time, moving down to the left of a node requires to decrease its *leftSize* field by *m*.

Let now *X* be the *i-th* node of the block that we have to delete, and let *n*=*m-i* be the number of nodes remaining to be removed after *X*.

If *n>0*, *before* removing X we must search and keep a reference to its inOrder successor Y. If X.right=NIL, and so we had to climb to find Y, Y.leftSize must be incremented by *n*.
For any value of *n*, if X has two children, during the descent to find the successor Y to be put in the place of $X^3$, we have to decrease by 1 the *leftSize* field of each visited node (Y excluded).

Once X has been deleted, it could also be necessary to manage rotations of possible fixes:
*Before* any rotation to the left, if the node to be rotated is Y, Y.right.leftSize must be decremented by *n*.
*Before* any rotation to the right, if Y is the left child of the node Z to be rotated, Z.leftSize must be incremented by *n*.

If *n>0*, we can now recursively proceed by deleting the already found node Y.

## Time complexity

The search for the nodes, as in a generic access, has an overall complexity of $T(n) = O(m + log(n))$.
So the cost is once again dominated by any possible restructuring procedure for every node, that in the worst-case correspond to $T(n) = O(m \cdot log(n))$[a][b], while in the average one $T(n) = O(m)$[b].

### 4.3.1 Technical explanation of the algorithm

Before starting, let me present the following statement:

**Theorem 4.3.1**
*Let X be a node of a (Wise) Red-Black Tree, such that X.left=NIL (X.right=NIL) and X.right≠NIL (X.left≠NIL).*
*Then, excluding sentinel leaves, the size of the subtree having X.right (X.left) as root is equal to 1.*

In other words, X.right (X.left) is the only node of the subtree.

**Proof**
*Consider the case where X.left=NIL and X.right≠NIL. The proof will obviously work even in the opposite case.*
*By definition of (Wise) Red-Black Tree, given a node X, the black heights of its two subtrees match. This means that, if X.left=NIL, the right subtree is fully composed by red nodes.*
*Again, by definition of (Wise) Red-Black Tree, we also have that a red node can only have black children. This means that the right subtree must only have one and one single (red) node.*

Let's move on. Managing the *leftSize* fields during the deletion of an entire block is not so easy.

---

[3] Obviously, during the node position exchange, the value of *leftSize* field of each node must not be changed, as well as all the other "structural" fields (*left*, *right*, *p*, *color*)

The trick is to focus on respecting a central rule:
*Before deleting a node X, the leftSize field of each node of the tree must correctly be setted considering that the whole block has already been deleted. This must be done in the hypothesis that all the remaining nodes of the block, from X forward, have X as ancestor.*
Note that, once we come up with the last item of the block, since there are no elements after it, every *leftSize* field will contain the correct value.

For the first node $X$ to be deleted, we just have to decrease by $m$ the *leftSize* field every time we move to the left during the descent.

Once $X$ is deleted, the rule should also apply for the successor $Y$, and obviously the successor of $X$ has to be searched *before* deleting $X$.
At first, let's assume the deletion of X will not require to restore any property.
If $X.right{=}NIL$, $Y$ is the closest ancestor of $X$ such that $X$ is in its left subtree, i.e. the deepest node previously decreased: we just have to increase $Y.leftSize$ by $n$.
If $X.right{\neq}NIL$ and $X.left{=}NIL$, for *Theorem 4.3.1* $X.right$ hasn't children: no operation is required.
If $X.right{\neq}NIL$ and $X.left{\neq}NIL$, this is the case where Y must take the place of $X$, and then be deleted. Again, the ancestors of $X$ will still have the correct *leftSize* value. We only have to decrease by 1 the field of the nodes we visit moving down from $X$ searching $Y$, in view of its following remotion (remember the $Y$ is the outermost left node of the subtree).

At this point, after deleting $X$, the rule will still work for Y.
However, we still have to consider possible fixes, and in particular rotations.
Let's consider first a left rotation of a node $Z$, and let $W$ be the right child of $Z$ *before* the rotation. As said previously, after the rotation $W$ will have $Z$ and the left subtree of $Z$ as additional nodes in its left subtree. If the next node $Y$ to be deleted is in this additional nodes, even $W.leftSize$ has to be preliminarily decremented. If $Y$ is in the left subtree of $Z$, the problem does not arise: the reason lies in the fact that $Z.leftSize$ is already decremented, and the rotation in Wise Red-Black Trees fix the *leftSize* value of $W$ based on the leftSize value of $W.left$. On the other hand, if $Y{=}Z$, we can just decrease $W.leftSize$ by $n$.
Consider now a right rotation of $Z$, with $W$ being the left child of $Z$ *before* the rotation.
This time, we know that $Z$ will lose $W$ and the left subtree of $W$. Once again, if $Y$ is in the left subtree of $W$, we don't need to do anything: in fact, both $Z$ and $W$ have been preliminarily decremented, and since the new value of $Z.leftSize$ is based on $Z.left.leftSize$, the only thing we care about is that the difference between $Z.leftSize$ and $W.leftSize$ is correct. If instead $Y{=}W$ we have to set $Z.leftSize$ to the correct value, incrementing it by $n$.

You can note that, since key values do not necessarily represent ranks, we are allowed to not shift any key of the nodes following the deleted ones.

# 5 Conclusion

To summarize, the search costs $T(n) = O(max(m, log(n)))$.

In the insertion procedure we don't have to do any search before adding a node, allowing us to calculate the cost based only on the rebalancing procedure. The first advantage is an average-case complexity that is only $T(n) = O(m)$ ($T(n) = O(max(m, log(n)))$ for a generic insertion). In the worst-case scenario, we still have a very good complexity of $T(n) = O(m \cdot log(n))$, and moreover we got rid of heavy multiplicative constants.

In the deletion, again, handling the block as a single unit is very useful in practical cases, reducing the non-visible factor to the asymptote from the worst-case complexity $T(n) = O(m \cdot log(n))$, and offering an average-case complexity of $T(n) = O(m)$.

All this was possible without sacrificing space, compared to using a list, or any other tree.